\documentclass[12pt,fleqn]{article}
\usepackage[utf8]{inputenc}
\usepackage[spanish, activeacute]{babel}
\usepackage{amsmath}
\usepackage{amssymb}
\usepackage{amsthm}
\usepackage{times}
\usepackage{mathptmx}
\usepackage{titletoc}
\usepackage{bm}
\usepackage{titlesec}
\usepackage[abbr]{harvard}
\usepackage{graphicx}
\usepackage{subfigure}
\usepackage{booktabs}
\usepackage{fancybox}
\usepackage{enumitem}
\setlist{nolistsep}
\usepackage{authblk}
\usepackage{booktabs}
\usepackage{multirow}

\usepackage{avant}
\usepackage{mathptmx}
\usepackage{microtype}
\usepackage{algorithm}
\usepackage{algpseudocode}
\usepackage{xcolor}
\numberwithin{equation}{section}
\usepackage[top=3cm,bottom=3cm,left=3.2cm,right=3.2cm,letterpaper]{geometry}
\titleformat*{\section}{\fontsize{12}{10}\selectfont\bfseries}
\titleformat*{\subsection}{\normalsize\bfseries}
\titleformat*{\subsubsection}{\normalsize\bfseries}
\newtheoremstyle{blacknumex}
{5pt} 
{5pt} 
{\normalfont} 
{} 
{\normalsize\bf} 
{\;\;} 
{0.1em} 
{\normalsize\thmname{#1}\thmnumber{{ }{#2}} 
\thmnote{{\bfseries--- #3.}}} 
\theoremstyle{blacknumex}
\newtheorem{theoremeT}{Teorema}[section]
\newtheorem{propositionT}{Proposición}[section]
\newtheorem{definitionT}{Definición}[section]
\newtheorem{corollaryT}{Corolario}[section]
\newtheorem{lemmaT}{Lema}[section]
\newtheorem{exampleT}{Ejemplo}[section]
\newtheorem{exerciceT}{Ejercicio}[section]
\newtheorem{remarkT}{Observación}[section]

\contentsmargin{0cm}
\titlecontents{section}[0.5cm]
{\addvspace{0pt}\normalsize}
{\contentslabel[\normalsize\thecontentslabel]{0.5cm}}
{}
{\normalsize\;\titlerule*[.5pc]{.}\;\thecontentspage}
[]
\titlecontents{subsection}[1.25cm]
{\addvspace{0pt}\normalsize}
{\contentslabel[\normalsize\thecontentslabel]{0.75cm}}
{}
{\normalsize\;\titlerule*[.5pc]{.}\;\thecontentspage}
[]
\titlecontents{subsubsection}[2.25cm]
{\addvspace{0pt}\normalsize}
{\contentslabel[\normalsize\thecontentslabel]{1.0cm}}
{}
{\normalsize\;\titlerule*[.5pc]{.}\;\thecontentspage}
[]

\author[,1]{Miguel S. Concha-Aracena\thanks{miguel.concha.21@alumnos.uda.cl}}%
\author[,1]{Leonardo Barrios-Blanco\thanks{leonardo.barrios.2020@alumnos.uda.cl}}%
\author[,1]{David Elal-Olivero\thanks{david.elal@uda.cl}}
\author[,2]{Paulo Henrique Silva
\thanks{paulohenri@ufba.br}}%
\author[,1]{Diego Carvalho do Nascimento\thanks{diego.nascimento@uda.cl}}%
\affil[1]{\fontsize{12}{10}\selectfont\textit{Departamento de Matemática, Universidad de Atacama, Copiapó, Chile}}
\affil[2]{\fontsize{12}{10}\selectfont\textit{Department of Statistics, Federal University of Bahia, Salvador, Brazil}}

\begin{document}
\title{\textbf{Extending normality: A case of unit distribution generated from the moments of the standard normal distribution}}
\date{}
\maketitle
\begin{abstract}
\noindent This article presents an important theorem, which shows that from the moments of the standard normal distribution one can generate density functions originating a family of models. Additionally, we discussed that different random variable domains are achieved with transformations. For instance, we adopted the moment of order two, from the proposed theorem, and transformed it, which allowed us to exemplify this class as unit distribution. We named it as Alpha-Unit (AU) distribution, which contains a single positive parameter $\alpha$ ($\text{AU}(\alpha) \in [0,1]$). We presented its properties and showed two estimation methods for the $\alpha$ parameter, the maximum likelihood estimator (MLE) and uniformly minimum-variance unbiased estimator (UMVUE) methods. In order to analyze the statistical consistency of the estimators, a Monte Carlo simulation study was carried out, where the robustness was demonstrated. As real-world application, we adopted two sets of unit data, the first regarding the dynamics of Chilean inflation in the post-military period, and the other regarding the daily maximum relative humidity of the air in the Atacama Desert. In both cases shown, the AU model is competitive, whenever the data present a range greater than 0.4 and extremely heavy asymmetric tail. We compared our model against other commonly used unit models, such as the beta, Kumaraswamy, logit-normal, simplex, unit-half-normal, and unit-Lindley distributions. 
\end{abstract}


\section{Introduction}
\label{sec:introduction}

Statistical methodology plays an important role in quantitative methods, given the hypothesis testing and inferential procedures. Nonetheless, the comparison across features is given based on a generated function estimated from the data information. Most often, mild suppositions are taken compromising the generalization of the results.

Under the perspective of statistical generalization (inferential method), some challenges are found for bounded distribution estimation. For instance, the confidence interval, which is often adopted from the maximum likelihood estimation approach and asymptotic supposition, is also assumed. Specially, interval estimation can be seen off the parameter space domain.

One exemplification is the case where bounded information data are observed, nonetheless, normality is commonly assumed to be true. This is the case of proportion/rate data, which are double bounded in the lower limit equal to 0 and upper limit equal to 1. Relative humidity is an example of this scenario where every decision-making should be $\in [0,1]$ \cite{fonseca2021water,bayer2018beta}, or commonly rates used in the field of finance, economics and demography, to list a few.

In the case of rates and proportions processes, as well as other processes whose variable of interest assumes values in the range $(0,1)$, there is a well-represented class of models, the unit distributions family, which deals with this type of double-bounded data. Among the existing unit distributions, we can cite the power distribution, beta distribution \cite{ferrari2004beta}, Kumaraswamy distribution \cite{kumaraswamy1980generalized}, unit-logistic distribution \cite{tadikamalla1982systems}, simplex distribution \cite{barndorff1991some}, unit-Weibull distribution \cite{mazucheli2018unitw,mazucheli2020unitw}, unit-Lindley distribution \cite{mazucheli2019ul}, unit-half normal distribution \cite{bakouch2021flexible}, unit log-log distribution \cite{korkmaz2021unitll}, modified Kumaraswamy and reflected modified Kumaraswamy distributions \cite{sagrillo2021modified}, unit-Teissier distribution \cite{krishna2022unitt}, unit extended Weibull families of distributions \cite{guerra2021unit}, unit folded normal distribution \cite{korkmaz2022unit}, unit-Chen distribution \cite{korkmaz2022unitc}, and Marshall-Olkin reduced Kies distribution \cite{afify2022new}.

Despite the applicability of the unit distributions in double-bounded variables, another important fact is that the interval estimation for the parameter may also be limited in a domain (like positive real number). In this manner, we also presented an inferential alternative through the delta method. 

This work starts by presenting an important theorem that transforms from a modification of the standard normal distribution into a class of density distributions that can be seen as unit. Then, as an exemplification, a case of second moment was chosen to illustrate the usefulness of this class of probabilistic models. This class of distributions shows to be competitive for high-frequency data with range greater than 0.4, important to real-world applications, whereas classical unit distribution fails \cite{santana2022beta}. Additionally, two different data sets were selected to illustrate the adjustment of the proposed model. The first is related to Chilean inflation (ultimate post-military era), and the second is from the dryest area of the planet (excluding the north and south poles).

\subsection{Motivation}

The normal distribution is very important in the history of statistics, where numerous modifications to this distribution have been proposed in the 
 literature \cite{stahl2006evolution,limpert2011problems}. An interesting fact related to the normal distribution is that its even moments can be used to generate new distributions, as is the case that we will show below, through a definition and a result embodied in a theorem that accounts for the characterization of these new distributions.

\medskip

\noindent \textbf{Definition 1.} 
A random variable $B$ is said to be distributed according to a Bimodal Normal (BN) distribution of order $k$, that is, $B \sim \text{BN}(k)$, discussed in \cite{elal2010alpha}, if its probability density function (PDF) is given by
\begin{equation}
f(b \mid k) = \frac{1}{c} b^{2k}\phi(b), \qquad b \in \mathbb{R},
\end{equation}
where $\phi(\cdot)$ is the PDF of a standard normal distribution, $c = \prod_{j=1}^{k} (2j-1)$ and $k=\{1,2,3,\ldots\}$.

\medskip 

\begin{figure}[ht]
    \centering
    \includegraphics[scale=0.45]{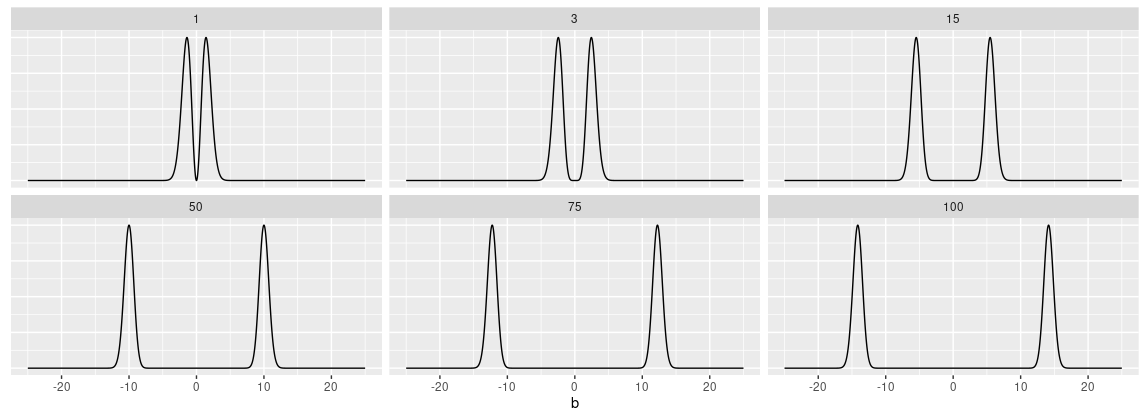}
    \caption{Density function of the BN distribution by varying the parameter $k$ (displayed on the top of each chart).}
    \label{fig:1}
\end{figure}

This class of distributions is always bimodal, where the observed modes move away when the order $k$ increases (see Figure \ref{fig:1}).

It is interesting to mention that transformations derived from the $\text{BN}(k)$ distribution may lead to other domains of interest, e.g., the unit domain. For example, let $B \sim \text{BN}(k)$, then by adding a scale parameter $\alpha$, the transformation $\alpha|B| \in \mathbb{R}^+$, and then the transformation $e^{-\alpha|B|} \in [0,1]$. Therefore, the stochastic characterization of a $\text{BN}(k)$ distribution can be obtained according to the following theorem.

\medskip

\noindent \textbf{Theorem 1.} 
Let $W_1$ and $W_2$ be independent random variables, where $W_1$ is such that $\mathbb{P}(W_1=1)=\mathbb{P}(W_1=-1) = 1/2$ and $W_2 \sim \chi^2_{2k+1}$. Then,
\begin{equation}\label{theom}
W_1 \sqrt{W_2} \sim \text{BN}(k).
\end{equation}

\medskip


So, this theorem is mainly motivated by the result that shows that if $X \sim \text{BN}(k)$, then $X^{2} \sim \chi^2_{2k+1}$. The demonstration is presented in Appendix A.

\section{Distribution of the Second Moment of the Unit-Normal Distribution}
\label{sec:bsb-distribution}
In this section, we will discuss a new unit distribution, named \textit{Alpha-Unit}, which presents a single parameter, $\alpha$. Whereas it will be presented its stochastic representations (probability density and cumulative distribution functions), moments, characteristic function, and how to generate random numbers from it. 


By taking the general theorem presented, and considering $k=1$, that is, considering the second moment of the standard normal distribution and its transform, a new unit distribution called Alpha-Unit will be illustrated. However, as $k$ increases, the concentration of the distribution intensifies.

\subsection{Properties and Characterization}

\noindent \textbf{Definition 2.} \textit{(Alpha-Unit distribution).} 
A random variable $X$ follows an Alpha-Unit (AU) distribution with parameter $\alpha>0 $, that is, $X \sim \text{AU}(\alpha)$, if its PDF is given by
\begin{equation}\label{eq:densidad-au}
f_{X}(x \mid \alpha)=\frac{2}{x \alpha} \left(\frac{\ln (x)}{\alpha}\right)^{2} \phi\left(\frac{\ln (x)}{\alpha}\right), \qquad 0< x \leq 1.
\end{equation}


\medskip


\noindent \textbf{Remark 1.} 
If $X \sim \text{AU}(\alpha)$, then its PDF is unimodal.

\medskip

\noindent \textbf{Demonstration.} 
The maxima of the AU distribution are studied, for which the criterion of the first derivative is first considered:
\begin{equation*}\label{}
\frac{d f_{X}(x \mid \alpha)}{dx} = \frac{2}{x \alpha^2} \frac{\ln(x)}{\alpha} \phi\left(\frac{\ln (x)}{\alpha}\right)   \left[ \frac{2}{x} -\frac{\ln(x)}{x} - \frac{[\ln(x)]^2}{\alpha} \frac{1}{x \alpha}  \right]=0.
\end{equation*}

Solving algebraically for $x$, we obtain:
\begin{equation*}
   x = 
   \begin{cases} 
      e^{ \left( \frac{-\alpha^2 +\sqrt{\alpha^4 + 8 \alpha^2}}{2} \right)}  & \mbox{(i) }   \\
      \\
      e^{ \left( -\frac{\alpha^2 +\sqrt{\alpha^4 + 8 \alpha^2}}{2} \right)}  & \mbox{(ii) } 
   \end{cases}.
\end{equation*}

To see if either or both expressions are solutions, it must be true that
\begin{equation*}
e^{-h} \ \ \& \ \ h>0.
\end{equation*}


By working algebraically, it can be seen that this is only true for (i), therefore, the AU distribution is unimodal. \quad $\blacksquare$


\medskip

\noindent \textbf{Proposition 1.} 
If $X \sim \text{AU}(\alpha)$, then its $r$-th order moment is given by
\begin{equation}
\mathbb{E}[X^{r}]= 2  e^{\left(\frac{r^{2} \alpha^{2}}{2}\right)} \left[\left(1+r^{2} \alpha^{2}\right)\left(1-\Phi(r \alpha)\right) - r \alpha \phi(r\alpha) \right].
\end{equation}

\medskip

\noindent \textbf{Demonstration.}
\begin{equation*}
\mathbb{E}[X^{r}]=  \int_{0}^{1} x^r \frac{2}{x \alpha} \left(\frac{\ln (x)}{\alpha}\right)^{2} \phi\left(\frac{\ln (x)}{\alpha}\right) dx. 
\end{equation*}
By making the change of variables:
\begin{equation*}
\begin{cases} u=\frac{1}{\alpha} \ln (x) \quad \Rightarrow \quad  e^{u \alpha}=x
\\
\\
du=\frac{1}{\alpha x} dx \quad \Rightarrow \quad \alpha e^{u \alpha} du =dx
\end{cases},
\end{equation*}
then substituting into the previous equation and developing algebraically, we have:
\begin{equation*}
\mathbb{E}[X^{r}]=2 e^{\frac{\alpha^2 r^2}{2}} \int_{-\infty}^{0}
u^2 \frac{1}{\sqrt{2 \pi}} e^{-\frac{(u-\alpha r)^2}{2}} du.
\end{equation*}

Then, by making another change of variables: $h=u-\alpha r$, $dh=du$; and replacing these expressions in the previous equation, we have:
\begin{align*}
\mathbb{E}[X^{r}] &= 2 e^{\frac{\alpha^2 r^2}{2}} \int_{-\infty}^{-\alpha r}
\left(h+\alpha r\right)^2 \frac{1}{\sqrt{2 \pi}} e^{-\frac{h^2}{2}} dh
\\
&= 2 e^{\frac{\alpha^2 r^2}{2}} \int_{-\infty}^{-\alpha r}
\left(h^2 +2h\alpha r+ \alpha^2 r^2\right) \phi(h)        dh
\\
&= 2 e^{\frac{\alpha^2 r^2}{2}} 
\left(
\int_{-\infty}^{-\alpha r}
h^2 \phi(h) dh+
2 \alpha r \int_{-\infty}^{-\alpha r}
h \phi(h) dh+
\alpha^2 r^2 \int_{-\infty}^{-\alpha r}
\phi(h) dh
\right).
\end{align*}

Solving the integrals, we get:
\begin{equation*}
\mathbb{E}[X^{r}]= 2 e^{\frac{\alpha^2 r^2}{2}} 
\left[
\alpha r \phi(\alpha r) 
+\left(1- \Phi(\alpha r) \right)
- 2 \alpha r \phi(\alpha r)
+ \alpha^2 r^2 \left(1- \Phi(\alpha r) \right)
\right].
\end{equation*}

Then, solving algebraically, we arrive at Proposition 1.  \quad $\blacksquare$

\medskip

From Proposition 1, we obtain the mean and variance of the $\text{AU}(\alpha)$ model as follows:
\begin{align*}
    \mathbb{E}[X]&=2e^{\frac{\alpha^2}{2}}\left[(1+\alpha^2)\left(1-\Phi(\alpha)\right)-\alpha \phi(\alpha)\right],\\
    \mathbb{V}\text{ar}[X]&=\mathbb{E}[X^2]-\left(\mathbb{E}[X]\right)^2\\ &=2e^{2\alpha^2}\left[(1+4\alpha^2)\left(1-\Phi(2\alpha)\right)-2\alpha \phi(2\alpha)\right]-4e^{\alpha^2}\left[(1+\alpha^2)\left(1-\Phi(\alpha)\right)-\alpha \phi(\alpha)\right]^2,
\end{align*}
where $\Phi(\cdot)$ is the cumulative distribution function (CDF) of a standard normal distribution.

\medskip

\begin{figure}[ht]
    \centering
    \includegraphics[scale=0.45]{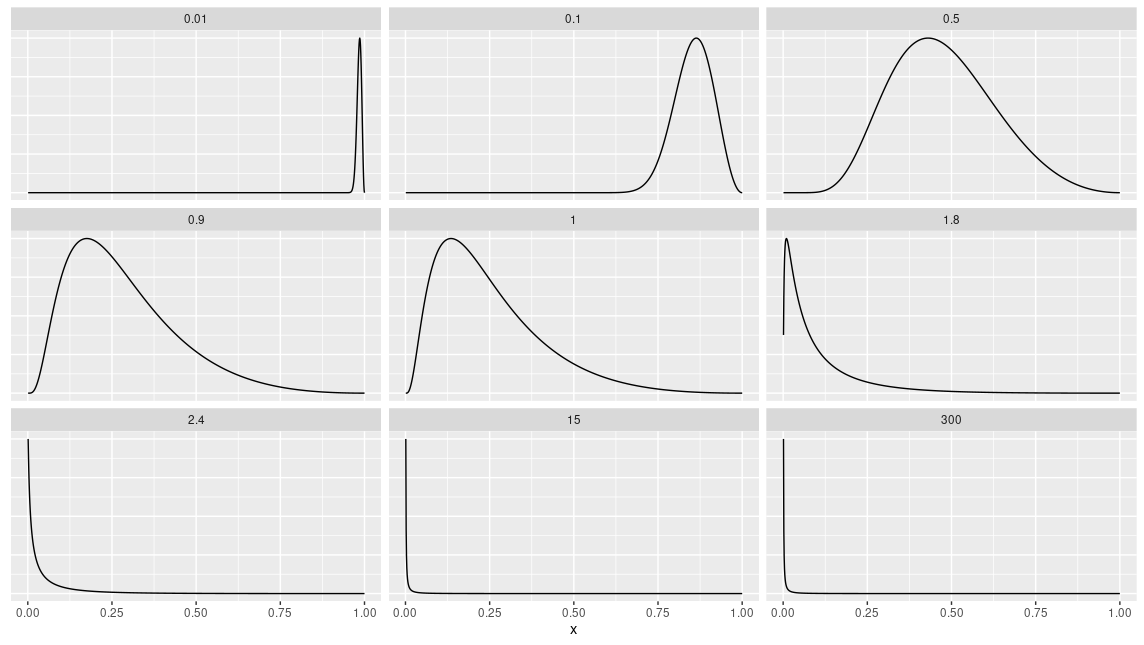}
    \caption{Density function of the AU distribution by varying the parameter $\alpha$ (displayed on the top of each chart). Whereas $B \sim \text{BN}(1) \rightarrow  B^2 \sim \chi^2_3$, then the AU model was generated from $X=e^{-\alpha|B|}$.}
    \label{fig:dNB}
\end{figure}

\noindent \textbf{Remark 2.} As illustration, Figure \ref{fig:dNB} shows the generated asymmetry and kurtosis based on the chosen $\alpha$ parameter of the AU distribution.

\noindent \textbf{Proposition 2.} 
If $X \sim \text{AU}(\alpha)$, then its CDF is given by
\begin{equation}\label{cdf_mi}
F_{X}(x \mid \alpha)= 2\Phi\left(\frac{\ln (x)}{\alpha}\right) - 2\left(\frac{\ln (x)}{\alpha}\right)\phi\left(\frac{\ln (x)}{\alpha}\right).
\end{equation}

\medskip


\noindent \textbf{Demonstration.} 
By definition of the CDF, we have:
\begin{equation*}
F_{X}(x \mid \alpha)= 
 \int_{0}^{x} \frac{2}{t \alpha} \left(\frac{\ln (t)}{\alpha}\right)^{2} \phi\left(\frac{\ln (t)}{\alpha}\right) dt. 
\end{equation*}

By making the change of variables:
\begin{equation*}
\begin{cases}
u=\frac{\ln (t)}{\alpha} \quad \Rightarrow \quad e^{u\alpha}=t 
\\
\\
du = \frac{1}{\alpha t} dt \quad \Rightarrow \quad \alpha e^{u \alpha} du = dt
\end{cases},
\end{equation*}
then substituting into the previous equation and reducing expressions algebraically, we get:
\begin{equation*}
F_{X}(x \mid \alpha)= 2 
\int_{- \infty}^{\frac{\ln (x)}{\alpha}} 
u^2 \phi (u) du.
\end{equation*}

Calculating the integral, we have:
\begin{align*}
F_{X}(x \mid \alpha) &= 2 \left[
-u \phi(u) \Big|_{-\infty}^{\ln (x)/\alpha}
+ \int_{-\infty}^{\ln (x)/\alpha} 
\phi (u) du 
\right]
\\
&= 2 \left[
-\left(\frac{\ln (x)}{\alpha}\right)\phi\left(\frac{\ln (x)}{\alpha}\right)
+\Phi\left(\frac{\ln (x)}{\alpha}\right) 
\right].
\end{align*}

Multiplying and commuting, we arrive at the expression of Proposition 2.  \quad $\blacksquare$

\medskip

Additionally, if $X$ denotes the monitored variable, then the PDF of $X$ is given by \eqref{eq:densidad-au}. Also, consider that the probability of false alarm (known as type I error) is $\pi$. Thus, we have:
\begin{equation*}
\mathbb{P}(X < \text{LCL} \mid \alpha) = \mathbb{P}(X > \text{UCL} \mid \alpha) = \pi / 2,
\end{equation*}
where $\alpha$ is the in-control process parameter (that is, the parameter that controls the quality characteristic based on the in-control state), LCL and UCL are the lower and upper control chart limits, respectively. Given the CDF $F_X(x \mid \alpha)$, then the quantile function of $X$ is defined by $Q(p \mid \alpha) = F_{X}^{-1}(p \mid \alpha)$, and can be 
obtained by setting to 0 and solving (numerically) for $x$ the following equation:
\begin{equation}\label{eq:Q}
    \Phi\left(\frac{\ln (x)}{\alpha}\right) - \left(\frac{\ln (x)}{\alpha}\right)\phi\left(\frac{\ln (x)}{\alpha}\right)- \frac{p}{2}, \qquad \text{for} \,\,\, 0 < p < 1.
 \end{equation} 

Following \cite{bayer2018CIE}, the control limits and centerline (CL) of the proposed AU control chart are given by
\begin{equation} \label{limits}
\text{LCL}  = Q\left(\pi/2 \mid \alpha\right), \qquad
\text{CL} = \mathbb{E}[X \mid \alpha], \\ \qquad
\text{UCL} = Q\left(1-\pi/2 \mid \alpha\right),
\end{equation}
where $Q(.)$ is the quantile function 
of the $\text{AU}(\alpha)$ distribution.

\medskip

\noindent \textbf{Proposition 3.} 
If $X \sim \text{AU}(\alpha)$, then its moment-generating function (MGF) is given by
\begin{equation}
\psi_{X}(t \mid \alpha)= 2 \sum_{k=0}^{\infty}\frac{t^{k}}{k!} e^{\left(\frac{k^{2} \alpha^{2}}{2}\right)} \left[\left(1+k^{2} \alpha^{2}\right)\left(1-\Phi(k \alpha)\right) - k \alpha \phi(k\alpha) \right].
\end{equation}

\medskip

\noindent \textbf{Demonstration.} 
By definition of the MGF, we have:
\begin{equation*}
\psi_{X}(t \mid \alpha)=
\mathbb{E} \left[e^{tx} \right] =
\int_{0}^{1} e^{tx} \frac{2}{x \alpha} \left(\frac{\ln (x)}{\alpha}\right)^{2} \phi\left(\frac{\ln (x)}{\alpha}\right) dx. 
\end{equation*}

By making the following change of variables:
\begin{equation*}
\begin{cases}
u= \frac{\ln (x)}{\alpha} \quad \Rightarrow \quad e^{u \alpha}=x 
\\
\\
du= \frac{1}{\alpha x} dx \quad \Rightarrow \quad \alpha e^{u \alpha} du = dx
\end{cases},
\end{equation*}
then substituting and simplifying into the previous equation, we get:
\begin{align*}
\psi_{X}(t \mid \alpha) &=
2 \int_{-\infty}^{0} e^{\left(t e^{u \alpha}\right)} u^2 \phi(u) du
\\
&=
2 \int_{-\infty}^{0} 
\sum_{k=0}^{\infty} \frac{t^k e^{u \alpha k} }{k!} u^2 \phi(u) du.
\end{align*}

Working algebraically, we have:
\begin{equation*}
\psi_{X}(t \mid \alpha)=
2 \sum_{k=0}^{\infty} \frac{t^k}{k!} e^{ \left(\frac{\alpha^2 k^2}{2} \right)}
\int_{-\infty}^{0} u^2 \frac{1}{\sqrt{2 \pi}}
e^{\left( \frac{-(u-\alpha k)^2}{2}  \right)} du.
\end{equation*}

By making the following change of variables: $h=u-\alpha k$, then substituting into the previous equation, we have: 
\begin{equation*}
\psi_{X}(t \mid \alpha)=
2 \sum_{k=0}^{\infty} \frac{t^k}{k!} e^{\left(\frac{\alpha^2 k^2}{2}\right)}
\int_{-\infty}^{-\alpha k} (h+\alpha k)^2 
\phi (h) dh. 
\end{equation*}

Then, solving the integral and adjusting algebraically, we arrive at the expression of Proposition 3. \quad $\blacksquare$

\medskip


\begin{algorithm}[ht]
    \caption{Random number generation from the $\text{AU}(\alpha)$ model.}
    \begin{algorithmic}
    \State \textbf{Step 1.} Generate a random number $X_1 \sim \chi^2_3$. \\
    \State \textbf{Step 2.} Generate a random number $U \sim \text{Uniform}(0,1)$. If $U \leq 1/2$, set $V=\sqrt{X_1}$; otherwise, $V=-\sqrt{X_1}$. \\
    \State \textbf{Step 3.} Based on the numbers obtained, generate the variable  $Y=\alpha|V|$, where $\alpha$ is a (positive) scale parameter and $|V|$ is a Bimodal Half-Normal (BHN).\\ 
    \State \textbf{Step 4.} Conclude with the number generated by Step 2 as a negative power of base $e$, that is, $X=e^{-Y}=e^{-\alpha|V|} \in [0,1]$. \\  
    \State \textbf{Step 5.} Repeat Steps 1-4 $n$ times to obtain a random sample of size $n$ from the $\text{AU}(\alpha)$ model.    
    \end{algorithmic}
\end{algorithm}

The pseudo-code describes the important steps for the generation of random numbers from the AU distribution. Further proofs are attached under Appendix B. 

\section{Inference}
\label{sec:inferential-aspect}
In this section, we will discuss the parameter estimation adopting the uniformly minimum-variance unbiased estimator (UMVUE) and maximum likelihood estimator (MLE) approaches. First, it will be demonstrated that the UMVUE is obtained straightforward since the proposed AU distribution is part of the exponential family. Later, the MLE will be also discussed, in which it will help to estimate not only with the point estimation of the $\alpha$ parameter, but also with the interval estimation. We enrolled the reasoning considering the asymptotic convergence in distribution of the parameter estimator, as well as we adapted a transformation which ensures that the interval of the parameter will always be on its domain (the delta method). The delta transformation procedure will enable the correct inferences and the standard error calculation associated with the parameter estimate. Later on, we will present a simulation study to illustrate these theoretical results.

\subsection{UMVUE through the Exponential Family}
Many of the distributions used in statistics belong to the exponential family, thereby implying a considerable advantage over other models that do not belong to this family. Such an advantage is declared significantly when it comes to calculating the statistic $T(\boldsymbol{X})$ of a random sample $\boldsymbol{X}=\left(X_{1},X_{2},\ldots,X_{n}\right)^{\top}$. We will show below that the proposed $\text{AU}(\alpha)$ model belongs to this family.

A random variable $X$ is said to belong to a \textit{one-parameter exponential family} if its associated PDF $f(\cdot \mid \theta)$ can be written in the form
\begin{equation*}
f(x \mid \theta)= \exp\left\{c(\theta) T(x) + d(\theta) + S(x)\right\}.
\end{equation*}

Let $X \sim \text{AU}(\alpha)$, then the PDF of $X$ can be written in exponential form as follows:
\begin{equation*}
f(x \mid \alpha)= \exp\left\{-\frac{1}{2\alpha^{2}} [\ln (x)]^{2} -3 \ln(\alpha) + \ln \left(\frac{[\ln (x)]^{2}}{x \sqrt{2 \pi}}\right)\right\}.
\end{equation*}
Then, $X$ belongs to a one-parameter exponential family if we define
$$
c(\alpha)= -\frac{1}{2\alpha^{2}}, \quad T(x) = [\ln (x)]^{2}, \quad d(\alpha) = -3 \ln (\alpha), \quad \text{and} \quad S(x) = \ln \left(\frac{[\ln (x)]^{2}}{x \sqrt{2 \pi}}\right).
$$
Let $\boldsymbol{x}=\left(x_{1},x_{2},\ldots,x_{n}\right)^{\top}$ be an observation from the random sample $\boldsymbol{X}=\left(X_{1},X_{2},\ldots,X_{n}\right)^{\top}$, with $X_{i} \sim \text{AU}(\alpha)$, for $i=1,2,\ldots,n$, then the joint PDF presented in exponential form would be given by
\begin{equation*}
f(\boldsymbol{x} \mid \alpha)= \exp\left\{-\frac{1}{2\alpha^{2}} \sum_{i=1}^{n} [\ln (x_{i})]^{2} -3n \ln(\alpha) + \sum_{i=1}^{n}\ln \left(\frac{[\ln (x_{i})]^{2}}{x_{i} \sqrt{2 \pi}}\right)\right\},
\end{equation*}
from which it can be concluded that the statistic $T(\boldsymbol{X}) =\sum_{i=1}^{n} [\ln (X_{i})]^{2} $ is sufficient and complete, once the AU model is part of the exponential family.

\medskip

\noindent \textbf{Proposition 4.} 
Let $\boldsymbol{X}=\left(X_{1},X_{2},\ldots,X_{n}\right)^{\top}$ be a random sample with $X_{i} \sim \text{AU}(\alpha)$, for $i=1,2,\ldots,n$, and $T(\boldsymbol{X}) =\sum_{i=1}^{n} [\ln (X_{i})]^{2} $, then
\begin{equation*}
W_{n} =  \frac{1}{\alpha^{2}} T(\boldsymbol{X}) \sim \chi_{3n}^{2}.
\end{equation*}

\medskip

\noindent \textbf{Demonstration.} 
If $D=\left[\frac{\ln(X)}{\alpha}\right]^2$, then $D \sim \chi^2_3$. Thus, $n$ independent and identically distributed samples of $D$ will have the sum of $n$ $\chi^2_3$, which will result in a $\chi^2$ distribution with degrees of freedom equal to $3n$, that is, $\chi^2_{3n}$, since
\begin{align*}
    F_D(d)&=\mathbb{P}(D \leq d)=\mathbb{P}\left( \left[\frac{\ln(X)}{\alpha}\right]^2 \leq d \right)=\mathbb{P}\left(-\sqrt{d} \leq \frac{\ln(X)}{\alpha} \leq \sqrt{d}\right)\\
    &=\mathbb{P}\left(-\alpha\sqrt{d} \leq \ln(X) \leq \alpha\sqrt{d}\right)=\mathbb{P}\left(\ln(X) \leq \alpha \sqrt{d}\right)-\mathbb{P}\left(\ln(X) \leq -\alpha \sqrt{d}\right)\\
    &=1-\mathbb{P}\left(\ln(X) \leq -\alpha \sqrt{d}\right)=1-\mathbb{P}\left(X \leq e^{-\alpha \sqrt{d} }\right)=1-F_X\left(e^{-\alpha \sqrt{d} }\right),
\end{align*}
so,
\begin{align*}
    f_D(d)&=\frac{\partial F_D(d)}{\partial d}=f_X\left(e^{-\alpha \sqrt{d} }\right) \left(e^{-\alpha \sqrt{d} }\right)\left(\frac{\alpha}{2\sqrt{d}}\right)\\
    &=\frac{2}{\alpha e^{-\alpha \sqrt{d}}} \left( \frac{-\alpha \sqrt{d}}{\alpha}\right)^2 \phi\left( \frac{-\alpha \sqrt{d}}{\alpha}\right) e^{-\alpha \sqrt{d}} \frac{\alpha}{2 \sqrt{d}}\\
    &= 
    \frac{1}{\sqrt{d}} \left(\sqrt{d}\right)^2 \frac{1}{\sqrt{2 \pi}} e^{- \frac{(\sqrt{d})^2}{2}}=\frac{1}{\sqrt{2 \pi}} d^{1/2} \exp\left(-d/2\right) \ \ \equiv \ \ \chi^2_3. \quad \blacksquare
\end{align*}

\medskip


\medskip

\noindent \textbf{Proposition 5.} 
Let $\boldsymbol{X}=\left(X_{1},X_{2},\ldots,X_{n}\right)^{\top}$ be a random sample with $X_{i} \sim \text{AU}(\alpha)$, for $i=1,2,\ldots,n$, and $T(\boldsymbol{X}) =\sum_{i=1}^{n} [\ln (X_{i})]^{2} $, then
\begin{equation*}
S(\boldsymbol{X}) =  \frac{\Gamma\left(\frac{3n}{2}\right)\sqrt{2}}{\Gamma\left(\frac{3n+1}{2}\right)} \sqrt{T(\boldsymbol{X})}
\end{equation*}
is an unbiased estimator of $\alpha$.

\medskip

\noindent \textbf{Demonstration.} 
Remembering that if $X \sim \text{Gamma}(a,b)$ distribution, then $\mathbb{E}[X^k]=\frac{\Gamma(a+b)}{b^k \Gamma(a)}$. Once the parameter $\alpha$ is observed to be squared, it will be necessary to release it to find an unbiased estimator. So, considering the random variable $W_n^{1/2}$ (with $W_n$ as defined in Proposition 4), then
$$\mathbb{E} \left[(W_n)^{1/2}\right] =\frac{\Gamma \left( \frac{3n}{2}+\frac{1}{2}\right)}{2^{1/2} \Gamma \left( \frac{3n}{2} \right)},$$
so,
\begin{align*}
\mathbb{E}\left[\left(\frac{1}{\alpha^2} T(\boldsymbol{X})\right)^{1/2}\right]
&=\frac{\Gamma \left( \frac{3n}{2}+\frac{1}{2}\right)}{2^{1/2} \Gamma \left( \frac{3n}{2} \right)} \\
\mathbb{E} \underbrace{\left[ \sqrt{T(\boldsymbol{X})}\frac{\Gamma \left( \frac{3n}{2} \right)\sqrt{2} }{\Gamma \left( \frac{3n}{2}+\frac{1}{2}\right)}\right]}_{S(\boldsymbol{X})}
&=\alpha. \quad \blacksquare 
\end{align*}


\medskip

\noindent \textbf{Remark 3.} 
Considering the two previous propositions and resorting to the Lehmann-Scheffé theorem, we can conclude that $S(\boldsymbol{X})$ is UMVUE for $\alpha$.

\subsection{Estimation using the Maximum Likelihood Method}
Let $\boldsymbol{x}=\left(x_{1},x_{2},\ldots,x_{n}\right)^{\top}$ be a realization of the random sample $\boldsymbol{X}=\left(X_{1},X_{2},\ldots,X_{n}\right)^{\top}$ taken from the 
 $\text{AU}(\alpha)$ model. Then, the log-likelihood function is given by
\begin{equation*}
\ell(\alpha) = \text{constant} 
            - 3n\ln(\alpha) 
            -\Sigma_{i=1}^{n} \ln (x_i)
            + 2\Sigma_{i=1}^{n} \ln (\ln (x_{i})) -\frac{1}{2 \alpha^{2}} \Sigma_{i=1}^{n}[\ln (x_{i})]^{2}. 
\end{equation*}

The MLE of $\alpha$, i.e., $\widehat{\alpha}$, is found by solving the following equation:
\begin{equation*}
\frac{d\ell(\alpha)}{d\alpha} = -\frac{3n}{\alpha} + \frac{1}{\alpha^{3}}\Sigma_{i=1}^{n}[\ln (x_{i})]^{2} =0,  
\end{equation*}
giving
\begin{equation*}
\widehat{\alpha } =\left \{\frac{1}{3n} \sum_{i=1}^{n} [\ln (x_{i})]^{2} \right\}^{1/2}.   
\end{equation*}
On the other hand, the second derivative of $\ell(\alpha)$ evaluated at $\alpha = \widehat{\alpha}$ is negative, concluding that $\widehat{\alpha}$ is MLE for $\alpha$.

It is known that, under certain regularity conditions,
\begin{equation*}
\sqrt{n}\left(\widehat{\alpha }- \alpha\right) \xrightarrow{\text{D}} \text{N}\left(0, I^{-1}(\alpha)\right),
\end{equation*}
where $I(\alpha)= -\mathbb{E}\left[\frac{d^{2}\ell(\alpha)}{d\alpha^{2}}\right] = \frac{6n}{\alpha^{2}}$.

A two-sided $100(1-\pi)\%$ confidence interval for $\alpha$ can be calculated by
\begin{equation}\label{eqINT}
    \left[\widehat{\alpha} - z_{1 - \pi/2}\,\sqrt{\mathbb{V}\text{ar}\left[\widehat{\alpha}\right]} , \ \widehat{\alpha} + z_{1 - \pi/2}\,\sqrt{\mathbb{V}\text{ar}\left[\widehat{\alpha}\right]}\right],
\end{equation}
where $z_{q}$ is the $q$-th percentile of the standard normal distribution. The variance of $\widehat{\alpha}$ can be approximated by the inverse of the observed Fisher information, as
\begin{equation}
   \mathbb{V}\text{ar}\left[\widehat{\alpha}\right] = I^{-1}\left(\widehat{\alpha}\right) = \frac{\widehat{\alpha}^{2}}{6n}.
\end{equation}

Since $\alpha$ is a positive value and we cannot guarantee that the lower limit of the interval \eqref{eqINT} is positive, we resort to the delta method to remedy such a situation. For this, we define the function $g: [0 , \infty) \rightarrow \mathbb{R}$ as $g(\alpha) = \ln (\alpha) $, and knowing that
$$
\sqrt{n}\left(g(\widehat{\alpha}) - g(\alpha)\right) \xrightarrow{\text{D}}  \text{N}\left(0, I^{-1}\left(\alpha\right) \left[
\frac{dg\left(\alpha\right)}{d\alpha}
\right]^{2}\right),
$$
we can then obtain an approximate two-sided $100(1-\pi)\%$ confidence interval for $\alpha$ by
\begin{equation}\label{deltaMethod}
     \left[\widehat{\alpha} \exp \left(-\frac{z_{1 - \pi/2}}{\sqrt{6n}}\right), \ \frac{\widehat{\alpha}}{\exp \left(-\frac{z_{1 - \pi/2}}{\sqrt{6n}}\right)}\right].
\end{equation}

\subsection{Simulation Study}

To illustrate the presented inferences for the estimation of the AU distribution, in this subsection we compared (via simulation study) the MLE versus the UMVUE. Moreover, we considered the scenarios where the parameter $\alpha=\{0.1,0.3,0.5,0.7,1.1,1.5\}$, considering sample sizes $n=\{100,200,500\}$, through the Monte Carlo method with N=1,000 repetitions. All this procedure took into account the random number generator for the $\text{AU}(\alpha)$ distribution presented in Algorithm 1. All analyses made in this work adopted the open-source R software \cite{Rcran}.

In order to compare the performance of the proposed estimators (MLE and UMVUE), since the true parameter value is known, we adopted the performance metrics bias and mean squared error (MSE), which are defined as follows:
\begin{equation*}
    \text{Bias}(\alpha)=\frac{1}{\text{N}}\sum_{i=1}^{\text{N}} (\widehat{\alpha}_{i}
    - \alpha)
\quad \text{and} \quad
    \text{MSE}(\alpha)=\frac{1}{\text{N}}\sum_{i=1}^{\text{N}} (\widehat{\alpha}_{i}-\alpha)^2, 
\end{equation*}
where $\widehat{\alpha}_{i}$ is the estimate for $\alpha$ in the $i$-th iteration (point estimation). Additionally, based on the asymptotic results presented in this work, we also calculated the confidence interval (CI) length adopting the delta method from Equation \eqref{deltaMethod} (interval estimation). That is, it analyzed the average of all the upper limits of the confidence interval, as well as the average of all the lower limits, and then calculated their difference.

Table \ref{TAB:sim} shows the obtained average estimates of the parameter $\alpha$, varying the sample size $n$, as well as the corresponding bias, MSE and CI length (this last only for MLE) results. 

\begin{table}[ht]
\begin{center}
\caption{Average estimates, bias and MSE for the MLE and UMVUE of the single parameter ($\alpha$) of the AU distribution, considering different sample sizes ($n$).}
\label{TAB:sim}
\begin{scriptsize}
\begin{tabular}{cccccccccc}
 \hline
& & \multicolumn{3}{c}{MLE} & & &\multicolumn{3}{c}{UMVUE}     \\ \cmidrule{3-6}  \cmidrule{8-10}
 $n$ &  $\alpha$  &   Estimate  & Bias & MSE&  CI Length & &  Estimate  & Bias & MSE \\ \hline
100 & 0.1  & 0.0998 & -0.0001 &1.6930e-05& 0.0160 & &0.0999&-8.2264e-05& 1.6165e-05
 \\   
 200 & &  0.0999 &  -9.8758e-05& 8.7306e-06 &  0.0113 & & 0.0999 &  -5.7124e-05 & 8.7314e-06 
 \\ 
  500 & & 0.0999  & -3.3400e-06 & 3.5542e-06&  0.0071 & &   0.1000 & 1.3327e-05 & 3.5555e-06 
 \\ 
  \hline
  100 & 0.3    &0.2996&-0.0003&0.0002& 0.0480 & & 0.2999&-8.0656e-05& 0.0002 
 \\   
 200 &   & 0.2997 & -0.0002 &7.8575e-05& 0.0339 & & 0.2998 &  -0.0001 &  7.8582e-05 
 \\ 
  500 &   & 0.2999 & -1.0020e-05 & 3.1987e-05&  0.0214 &  &  0.3002 & 0.0002 &  3.0979e-05
 \\ 
  \hline
  100 & 0.5  & 0.4994 &-0.0005&0.0004 & 0.0800 & & 0.4999 & -0.0001 & 0.0004
 \\   
 200 &   & 0.4997 &-0.0004&0.0002& 0.0565 & &  0.4997 & -0.0003 &0.0002  
 \\ 
  500 &   & 0.4999 &-1.6700e-05&8.8855e-05& 0.0357 & &0.5000 & 6.6637e-05 & 8.8888e-05 
 \\ 
  \hline
  100 & 0.7  & 0.6992 & -0.0007 & 0.0008 & 0.1120 & &  0.6998 & -0.0002 & 0.0008   
 \\   
 200 &   & 0.6993 & -0.0006&0.0004& 0.0791 & &  0.6996 & -0.0004 & 0.0004 
 \\ 
  500 &   & 0.6999 &-2.3380e-05&0.0001&  0.0501 & &  0.7000 & 9.3291e-05 & 0.0001 
 \\ 
  \hline
  100 & 1.1  &  1.0987& -0.0012&0.0020& 0.1760 && 1.0997 & -0.0003 & 0.0020 
 \\   
 200 &   & 1.0989 & -0.0010&0.0010&  0.1244 &&  1.0994 & -0.0006 & 0.0010 
 \\ 
  500 &   & 1.0999 & -3.6741e-05& 0.0004 & 0.0787 &&  1.1001 & 0.0001 & 0.0004  
 \\ 
  \hline
  100 & 1.5  & 1.4983 &-0.0016&0.0038& 0.2400 &&  1.4996 &-0.0004 & 0.0038  
 \\   
 200 &   & 1.4985 &-0.0014&0.0019& 0.1696 &&  1.4991 & -0.0008 & 0.0019 
 \\ 
  500 &   & 1.4999 &-5.0101e-05&0.0008 & 0.1073 &&  1.5002 & 0.0002 &  0.0007 
 \\ 
 \hline
    \end{tabular}
    \end{scriptsize}
\end{center}
\end{table}

The asymptotic convergence of the MLE, towards the robustness, was noticed as the sample size increases. In addition, both MLE and UMVUE's bias and MSE are small and tend to decrease as $n$ gets larger. On the other hand, the CI length also decreases as the sample size increases.


As a last summary, regarding the robustness of the estimators, it was taken the difference between the MLE and UMVUE estimates, considering each different sample size $n$, then the interquartile range (IQR) was calculated per sample size group. That is, $\text{IQR}^{(ni)}(\widehat{\alpha_1}^{(ni)}_{\text{MLE}} - \widehat{\alpha_1}^{(ni)}_{\text{UMVUE}},\ldots,\widehat{\alpha_j}^{(ni)}_{\text{MLE}} - \widehat{\alpha_j}^{(ni)}_{\text{UMVUE}})$, where $ni=\{100,200,500\}$ and $\alpha_j=\{\alpha_1=0.1,\alpha_2=0.3,\ldots,\alpha_6=1.5\}$. For instance, the IQR for $n=100$ was $0.00053$, when for $n=200$ reduced to $0.00025$, and $n=500$ resulted in $0.00012$. This shows, in summary, that every time the sample size is large, the error range gets smaller, regardless the value of the $\alpha$ parameter.

\section{Real-World Exemplifications}

In this section, we exemplified two applications adopting the AU distribution with real-world problems. The first case is with respect to the dynamics of Chilean inflation in the post-military dictatorship period. And the second case is regarding the relative humidity of the air in the northern Chilean city of Copiapó (Atacama region). 

The Chilean inflation data are recorded annually, whose values considered range from 1992 to 2021. These are based on the period after the military dictatorship of 1973-1990. It was analyzed the dynamic of the inflation data (in \%), which were standardized by min-max transformation, resulting in a unit response variable (value between 0 and 1). The years 1990 and 1991 were excluded, since they are considered as a period of transition. Then, total amount of observations were 30 years (from 1992 until 2021).

On the other hand, the relative air humidity data cover a period from February 2015 to October 2022, with a one-hour recording format (104,415 observations). Then, this dataset was transformed into daily maximum observation (6,226 observations).

\subsection{Chilean Inflation (Post-Military Era)}

\begin{figure}[ht]
    \centering
    \includegraphics[scale=0.9]{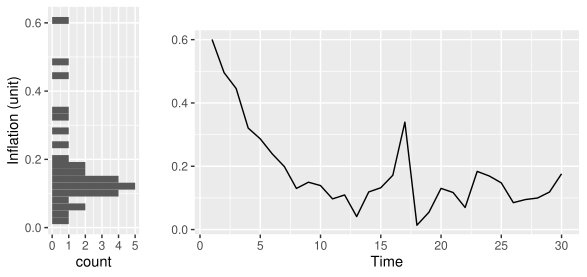} 
    \caption{Chilean inflation in the period 1992-2021 (post-military era). The histogram on the left shows a skewness of the data. The dynamic is represented in the right panel, where a disturbance (outlier) is noticed in the year 2008 (observation \#17).}
    \label{fig:Inflat}
\end{figure}

Figure \ref{fig:Inflat} presents the dynamics of Chilean inflation in the post-military dictatorship period, demonstrating stability between the years 1999 and 2008. The right panel shows the time series for inflation, where time is measured in years, from year 1 (1992) until year 30 (2021). The left panel shows the accumulation of the values of the time series, where a predominant trend is shown around 0.1 of the inflation rate.

\begin{figure}[ht]
    \centering
    \includegraphics[scale=0.43]{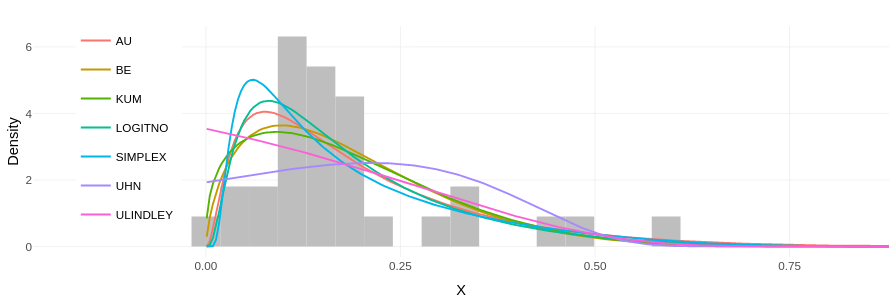} \qquad
    \includegraphics[scale=0.45]{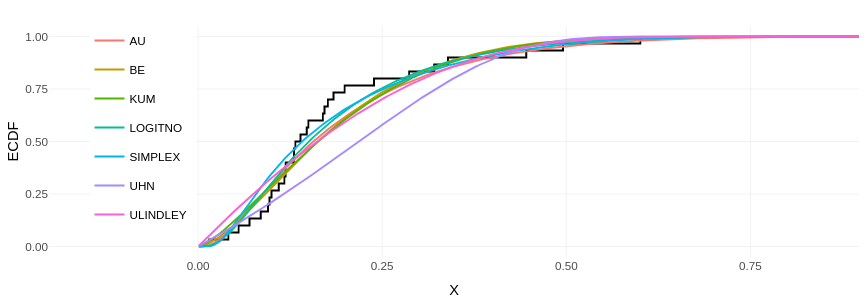}
    \caption{Estimated densities superimposed to the histogram (top-chart), and estimated CDFs superimposed to the ECDF (bottom-chart) (Chilean inflation data).}
    \label{fitInfla}
\end{figure}

Once the empirical dynamic of these data was analyzed, the most common unit distributions, presented in the statistical literature, were adjusted. The upper panel of Figure \ref{fitInfla} illustrates the histogram for the inflation data, where it is compared with different adjusted densities based on the MLE: AU, beta (BE), Kumaraswamy (KUM), logit-normal (LOGITNO), simplex (SIMPLEX), unit-half-normal (UHN), and unit-Lindley (ULINDLEY). The lower panel of the same figure shows the fitted CDFs superimposed to the empirical CDF (ECDF).

In order to quantify the performance of the adjusted models, we analyzed the Akaike Information Criterion (AIC) \cite{akaike1977}, and the Bayesian (or Schwarz) Information Criterion (BIC) \cite{schwarz1978}. The obtained results (see Table \ref{tabapli1}) show the AU model as the best-fitted model to this data set. In addition, it is possible to infer about the average of the phenomenon, that is, the expectation of the model AU($\widehat{\alpha}=1.2059$), resulting in $\mathbb{E}[X_\text{Inflation}]=0.1948$. In other words, the average Chilean inflation, post-military era, is 19.48\%. 

\begin{table}[ht]
    \centering
    \caption{Parameter estimates, AIC and BIC values (Chilean inflation data). S.E. $=$ standard error.}
    \begin{tabular}{llrr}
    \hline
        Model & Parameter Estimate (S.E.)  & AIC & BIC \\
         \hline
        $\text{AU}(\alpha)$ & $\widehat{\alpha}= 1.205943 \, (0.008079)$ & $-47.89$ & $-46.49$ \\
       
        $\text{BE}(\mu,\sigma)$ & $\widehat{\mu}= 0.185857 \, (0.000496)$ & $-44.58$ & $-41.78$ \\
         ~ & $\widehat{\sigma}= 0.314688 \, (0.001304)$ & ~ & ~ \\
       
       $\text{KUM}(\mu,\sigma)$ & $\widehat{\mu}= 1.370127 \,(0.045522)$ & $-43.63$ & $-40.83$ \\
         ~ & $\widehat{\sigma}= 7.968427 \, (7.750459)$ & ~ & ~ \\
      
         $\text{LOGITNO}(\mu,\sigma)$ & $\widehat{\mu}= 0.150323 \,(0.000457)$ & $-46.23$ & $-43.43$ \\
         ~ & $\widehat{\sigma}= 0.916938\, (0.014013)$ & ~ & ~ \\
        
      $\text{SIMPLEX}(\mu,\sigma)$  & $\widehat{\mu}=0.182462 \,(0.000584)$  & $-43.17$ & $-40.37$ \\
       ~ & $\widehat{\sigma}= 2.854833 \,(0.135834)$ & ~ & ~ \\
      
      $\text{UHN}(\sigma)$  & $\widehat{\sigma}=0.413894 \,(0.002855)$  & $-33.62$ & $-32.22$ \\
      
      $\text{ULINDLEY}(\mu)$  & $\widehat{\mu}=0.186834 \,(0.000575)$  & $-41.99$ & $-40.58$ \\
          \hline
    \end{tabular}
    \label{tabapli1}
\end{table}

In the following subsection, we will illustrate the performance of the AU model adopting a high-frequency data set originated from the relative humidity from a city located in the Atacama Desert. 

\subsection{Water Monitoring in Air Humidity}

The hydrological regime of the main rivers of Atacama is characterized by ice sources: water flows from the peaks following the melting of snowfall, glaciers, and permafrost located in the upper parts of the Andes range. In the context of climate change, it is therefore essential to understand the hydrological cycle of these regions in order to set up a sustainable management policy. Understanding the hydrological cycle requires the implementation of tools for forecasting river flows, relative humidity, groundwater reservoirs, or any other water-related quantity monitoring, which inevitably needs an in-depth knowledge of the physical phenomena that govern the entire hydrological cycle and, more precisely, the complex interaction between atmosphere, climate, landforms, ice, snow and river flows. 

\begin{figure}
    \centering
    \includegraphics[scale=0.63]{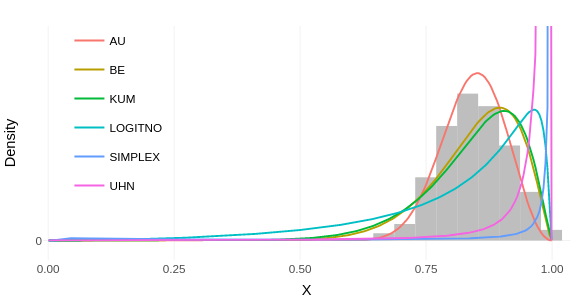} \qquad
    \includegraphics[scale=0.65]{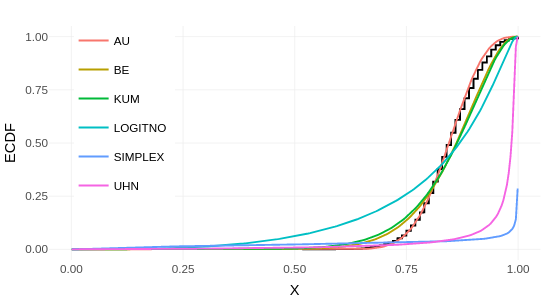}
    \caption{Estimated densities superimposed to the histogram (top-chart), and estimated CDFs superimposed to the ECDF (bottom-chart) (relative air humidity data).}
    \label{acumapli2}
\end{figure}

Additionally, a unique phenomenon called \textit{Camanchaca} happens, which is a fog passing by the Copiapó city, recurrent only between midnight until around 10 a.m. Here, we demonstrate the variation of the relative humidity of Copiapó city, proposing a methodology that can be efficient, adjustable to these data. Using the daily maximum relative humidity, we compared six different unit distributions: AU, BE, KUM, LOGITNO, SIMPLEX, and UHN, as shown in Figure \ref{acumapli2}.

After comparing the commonly used unit models, we demonstrate the advantage of adjusting the AU model against others (visually). Table \ref{tabapli2} confirms the best fit of the AU model based on information criteria (AIC and BIC), as well as shows the estimation of the parameter(s) of each model.

\begin{table}[ht]
    \centering
    \caption{Parameter estimates, AIC and BIC values (relative air humidity data).}
    \begin{tabular}{llrr}
    \hline
        Model & Parameter Estimate (S.E.)  & AIC & BIC \\
         \hline
        $\text{AU}(\alpha)$ & $\widehat{\alpha}=0.1092 \, (3.1902\text{e-}07)$  &$-\text{14,023.49}$	  & $-\text{14,016.76}$ \\
       
        $\text{BE}(\mu,\sigma)$ & $\widehat{\mu}=0.8476 \, (1.2027\text{e-}06) $ & $-\text{13,927.89}	$ & $-\text{13,914.41}$ \\
         ~ & $\widehat{\sigma}= 0.2410 \, (4.1119\text{e-}06)$ & ~ & ~ \\
       
       $\text{KUM}(\mu,\sigma)$ & $\widehat{\mu}= 9.4004 \,  (0.0141)$ & $-\text{13,605.90}$ & $	-\text{13,592.43}$ \\
         ~ & $\widehat{\sigma}= 2.3882 \,  (0.0019)$  & ~ & ~ \\
      
         $\text{LOGITNO}(\mu,\sigma)$ & $\widehat{\mu}=0.8693 \, (3.1376\text{e-}06)$ & $-\text{7,600.43}	$ & $-\text{7,586.95}$ \\
         ~ & $\widehat{\sigma}=1.2299 \, (1.2148\text{e-}04) $ & ~ & ~ \\
        
      $\text{SIMPLEX}(\mu,\sigma)$  & $\widehat{\mu}=0.9735 \, (1.2959\text{e-}06)$  & $\text{32,477.13}	$ & $\text{32,490.61}$ \\
       ~ & $\widehat{\sigma}= 94.0480 \, (0.7103)$ & ~ & ~ \\
      
      $\text{UHN}(\sigma)$  & $\widehat{\sigma}=99.9900 \, (6.5334\text{e-}07)$  & $\text{5,101,018,733.13}	$ & $\text{5,101,018,739.86}$ \\
      \hline
    \end{tabular}
    \label{tabapli2}
\end{table}

\begin{figure*}
    \centering
    \includegraphics[width=\textwidth]{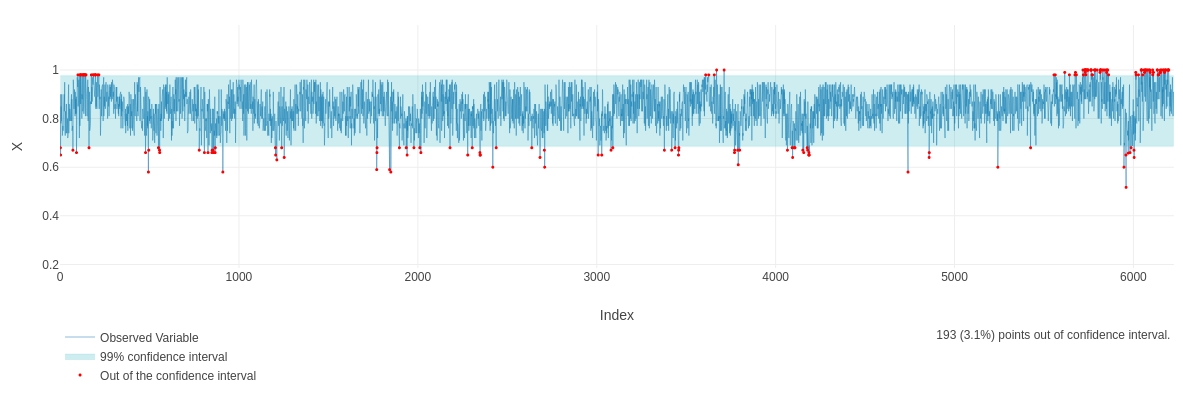}
    \caption{SPC control chart, considering a 99\% of tolerance based on the AU model fitted to the daily maximum relative humidity of Copiapó city, Chile, from February 1st, 2015 until October 4th, 2022. It is observed that 193 days (3.1\%) presented anomaly values (out-of-control signals). The obtained control limits were: $\text{LCL}=68.56\%$ and $\text{UCL}=97.73\%$.}
    \label{fig:SPC_water}
\end{figure*}

After obtaining the parameter estimate for $\alpha$, the AU model was used to construct a Statistical Process Control (SPC) chart, by calculating a tolerance upper-lower bound. Moreover, it was adopted the Highest Density Interval (HDI), considering a confidence of 99\% to monitor the daily maximum relative humidity records (as showed by Figure \ref{fig:SPC_water}).

The expected daily maximum water relative humidity is 76.48\% (based on the adjusted AU model). The obtained control limits, considering a confidence (or tolerance) of 99\%, were: $\text{LCL}=68.56\%$ and $\text{UCL}=97.73\%$. Thus, the control chart based on the AU model, or simply, AU control chart, is another exciting and valuable alternative to some well-known SPC tools, which enlightens the prediction and opens new doors to discuss extreme events in the Atacama water particles monitoring through probabilistic reasoning.

\section{Conclusions}
This work showed the competitiveness of the developed Theorem 1 (equation \eqref{theom}), which enables a great class of distributions that are all from the exponential family. As an exemplification, we adopted the special case for $k=1$, equivalent to the moment of order two of the standard normal distribution, and after some transformations, developed the Alpha-Unit (AU) distribution. Whereas we dedicated to the unit range given the importance of this stochasticity representation. 


Unit distributions are useful for values that oscillate between 0 and 1, such as fractions, proportions and rates, among others, or for a set of values where there is a minimum or maximum limitation, resorting to standardization through the min-max transformation. Most distributions of this type come from transforming a random variable with certain distribution so that it takes values between 0 and 1, as in the case of unit-Lindley distribution \cite{mazucheli2019ul}, which comes from the Lindley distribution \cite{lindley1958,lindley1965}.


There are numerous works based on (unit) distributions, extending a model and applying it to several areas \cite{korkmaz2021unitll,guerra2021unit,korkmaz2022unit}. In this work, we introduced and showed the competitiveness of the AU distribution, especially for data with a range greater than 0.4, or which present high asymmetry and low decay. Further studies shall investigate this hypothesis in a wider amount of data sets (through different sorts of wide data range). Additionally, implementation in this model adopting hierarchical estimation and spatio-temporal dependence would be useful for forecast/predictable problems.

\bibliographystyle{agsm}

\newpage
\section*{Appendix A}
This appendix shows the proof that for a random variable
\begin{equation*}
    X \sim \text{BN}(k) \quad \rightarrow \quad X^2 \sim \chi^2_{2k+1}.
\end{equation*}

Then,
\begin{align*}
    F_{X^2}(x)&=\mathbb{P}\left(X^2 \leq x\right)=\mathbb{P}\left(-\sqrt{x}\leq X \leq \sqrt{x}\right)=2\mathbb{P}\left(X\leq \sqrt{x}\right)-1=2F_{X}\left(\sqrt{x}\right)-1.
\end{align*}

It follows that
\begin{align*}
    f_{X^2}(x)&=2f_X\left(\sqrt{x}\right)\frac{1}{2 \sqrt{x}}=\frac{1}{c}\left(\sqrt{x}\right)^{2k}\phi(\sqrt{x})\frac{1}{\sqrt{x}}=\frac{1}{\prod^k_{j=1}(2j-1)}\left(\sqrt{x}\right)^{2k-1}\frac{1}{\sqrt{2\pi}}e^{-\frac{x}{2}}.
\end{align*}

Knowing that $\Gamma\left(\frac{2k+1}{2}\right)=\prod^k_{j=1}(2j-1) \frac{\sqrt{\pi}}{2^k}$, then
\begin{align*}   
   f_{X^2}(x)&=\frac{1}{\prod^k_{j=1}(2j-1)}x^{\frac{2k-1}{2} }\frac{1}{\sqrt{2\pi}}e^{-\frac{x}{2}}=\frac{\sqrt{\pi}}{2^k \Gamma\left(\frac{2k+1}{2}\right)}\frac{x^{\frac{2k-1}{2}}}{2^{1/2}\sqrt{\pi}}e^{-\frac{x}{2}}\\
    &=\frac{1}{ \Gamma\left(\frac{2k+1}{2}\right)2^{\frac{2k+1}{2}}}
    x^{\frac{2k-1}{2}}e^{-\frac{x}{2}}.
\end{align*}

Therefore, $X^2 \sim \chi^2_{2k+1}$.

Besides that, complementation can be taken into account by saying that, considering $W_2 \sim \chi^2_{2k+1}$ and $\mathbb{P}(W_1=\pm 1)=1/2$, then $B=W_1\sqrt{W_2} \sim \text{BN}(k)$:

Let $b \geq 0$, then
\begin{align*}
    F_B(b)&=\mathbb{P}(B \leq b)=\mathbb{P}\left(W_1\sqrt{W_2}\leq b\right)\\
    &=\mathbb{P}\left(W_1\sqrt{W_2}\leq b \mid W_1=1\right)\mathbb{P}\left(W_1=1\right)+\mathbb{P}\left(W_1\sqrt{W_2}\leq b \mid W_1=-1\right)\mathbb{P}\left(W_1=-1\right)\\
    &\overset{\text{ind.}}{=}\mathbb{P}\left((1)\sqrt{W_2}\leq b\right) \frac{1}{2}+\mathbb{P}\left((-1)\sqrt{W_2}\leq b\right) \frac{1}{2}.\\
    &\text{Since }b \geq0, \ \ \text{then} \ \ \mathbb{P}\left((-1)\sqrt{W_2}\leq b\right)=1: \\
    &=\mathbb{P}\left(\sqrt{W_2}\leq b\right) \frac{1}{2}+\frac{1}{2}=\mathbb{P}\left(|W_2|\leq b^2\right) \frac{1}{2}+\frac{1}{2}=\mathbb{P}\left(-b^2 \leq W_2 \leq b^2\right) \frac{1}{2}+\frac{1}{2}\\
    &=\frac{1}{2}\bigg[\mathbb{P}\left(X\leq b^2\right)-\underbrace{\mathbb{P}\left(X \leq -b^2\right)}_{0}\bigg]+\frac{1}{2}=\frac{1}{2}\mathbb{P}\left(X\leq b^2\right)+\frac{1}{2}=\frac{1}{2}F_X\left(b^2\right)+\frac{1}{2}\\
    &\text{Therefore, }\\
    f_B(b)&=\frac{\partial F_B(b)}{\partial b}=\frac{1}{2}f_X\left(b^2\right)2b=b f_X\left(b^2\right)=b \frac{1}{\Gamma\left(\frac{2k+1}{2}\right)2^{\frac{2k+1}{2}}}\left(b^2\right)^{\frac{2k+1}{2}-1}e^{-\frac{b^2}{2}}\\
    &=b \frac{1}{\Gamma\left(\frac{2k+1}{2}\right)2^{\frac{2k+1}{2}}}b^{2k-1}e^{-\frac{b^2}{2}}= \frac{1}{ \frac{\sqrt{\pi}\prod^k_{j=1}(2j-1)}{2^k} 2^{\frac{2k+1}{2}}}b^{2k}e^{-\frac{b^2}{2}}\\
    &=
    \frac{1}{\prod^k_{j=1}(2j-1)} \frac{b^{2k}}{\sqrt{2\pi}}e^{-\frac{b^2}{2}}=\frac{1}{\underbrace{\prod^k_{j=1}(2j-1)}_{c} } b^{2k} \phi(b).
\end{align*}

Analogously, it is proved for $b<0$.

\newpage
\section*{Appendix B}

The proposed theorem will be illustrated considering $k=1$ to show the origin of the random numbers that generate the AU distribution.

\medskip

\noindent \textbf{Proposition 6.} If $X \sim \text{BN}(1)$, then
\begin{enumerate}
  \item $f_{X}(x) = x^{2}\phi(x)$ is a bimodal density function;
  \item $X^{2} \sim \chi_{3}^{2}$;
  \item Let $Y$ and $V$ be independent random variables with $Y \sim \chi_{3}^{2}$ and $V$ is such that $\mathbb{P}(V=-1)=\mathbb{P}(V=1)= 1/2$, then
$$
R=V \sqrt{Y} \sim \text{BN}(1).
$$
\end{enumerate}


\noindent \textbf{Demonstration.}

\begin{enumerate}
\item 
If $f_X(x)$ is bimodal, it would have two maxima, for which the first and second derivative criteria would be applied:
\begin{equation*}
\frac{d f_X(x)}{dx}=0 \quad \Rightarrow
\end{equation*}
\begin{equation*}
\frac{d \left(x^2 \phi(x)\right)}{dx}= 2x \phi(x) +x^2 [-x \phi(x)]=2x \phi(x) -x^3 \phi(x)=x \phi(x) (2-x^2)=0.
\end{equation*}

Then, it can be seen that the solutions of the previous equation would be: $x_1=0$, $x_2=\sqrt{2}$, $x_3=-\sqrt{2}$. Hence, applying the second derivative criterion:
\begin{equation*}
\frac{d^2 f_X(x)}{dx^2}<0 \quad \Rightarrow
\end{equation*}
\begin{equation*}
\frac{d \left(x \phi(x) (2-x^2)\right)}{dx}= \phi(x) (2-x^2)+ x [-x \phi(x)] (2-x^2) + x \phi(x) (-2x).
\end{equation*}

Reducing algebraically, we get:
\begin{equation*}
\frac{d^2 f_X(x)}{dx^2}=\phi(x) \left(2-5 x^2 +x^4\right) <0.
\end{equation*}

The only solutions that satisfy the previous inequality are: $x_2=\sqrt{2}$, $x_3=-\sqrt{2}$. Therefore, there are two maxima and the BN distribution is bimodal.


\medskip

\item Let $W=X^2$ and $w>0$:
\begin{align*}
F_W(w)&=\mathbb{P}(W \leq w)= \mathbb{P}(X^2 \leq w)= \mathbb{P}(-\sqrt{w} \leq X \leq \sqrt{w}) \\ &= \mathbb{P}(X \leq \sqrt{w})- \mathbb{P}(X \leq -\sqrt{w}) \\
&=F_X \left(\sqrt{w}\right) - \left[1-F_X \left(\sqrt{w}\right)\right]= 2 F_X \left(\sqrt{w}\right)-1.
\end{align*}

Then, by deriving the previous expression, we obtain:
\begin{equation*}
f_W (w)= 2 f_X \left(\sqrt{w}\right) \frac{1}{2 \sqrt{w}}= \frac{1}{\sqrt{w}} f_X \left(\sqrt{w}\right)= \frac{1}{\sqrt{w}} \left(\sqrt{w}\right)^2 \phi\left(\sqrt{w}\right) = \frac{1}{\sqrt{2 \pi}} w^{1/2} e^{-w/2}.
\end{equation*}

Observing the expression above, we have that $X^{2} \sim \chi_{3}^{2}$. $\quad \blacksquare$ 
\end{enumerate}


\medskip

\noindent \textbf{Definition 3.} \textit{(Bimodal Half-Normal distribution).} 
Let $Y \sim \text{BN}(1)$, if $Q = \alpha |Y|$ with $\alpha > 0 $, then we say that $Q$ is distributed according to a Bimodal Half-Normal (BHN) distribution with parameter $\alpha $ and we denote it by $Q \sim \text{BHN}(\alpha)$.

\medskip

\noindent \textbf{Proposition 7.} 
If $Q \sim \text{BHN}(\alpha)$, then the PDF of $Q$ is given by
\begin{equation*}
f_{Q}(q \mid \alpha)= \frac{2}{ \alpha} \left(\frac{ q}{\alpha}\right)^{2} \phi\left(\frac{ q}{\alpha}\right), \quad q > 0. \\
\end{equation*}


\noindent \textbf{Demonstration.}
\begin{equation*}
F_Q (q)=\mathbb{P}(Q \leq q)= \mathbb{P}(\alpha |Y| \leq q)= \mathbb{P}\left(-\frac{q}{\alpha} \leq Y \leq \frac{q}{\alpha}\right) =2 \mathbb{P}\left(Y \leq \frac{q}{\alpha}\right)-1=2 F_Y \left(\frac{q}{\alpha}\right)-1.
\end{equation*}

Then, by deriving the previous expression, we obtain:
\begin{equation*}
f_Q(q)= 2 f_Y\left(\frac{q}{\alpha}\right) \frac{1}{\alpha}= \frac{2}{\alpha} \left(\frac{q}{\alpha}\right)^2 \phi\left(\frac{q}{\alpha}\right).  \quad \blacksquare
\end{equation*}

\medskip

\noindent \textbf{Proposition 8.} If $Q \sim \text{BHN}(\alpha)$, then
\begin{equation*}
X = e^{-Q} \sim \text{AU}(\alpha). \\
\end{equation*}

\medskip

\noindent \textbf{Demonstration.} Let $X=e^{-Q}$, $0 < x\leq 1$, then
\begin{align*}
F_X(x)&=\mathbb{P}(X\leq x)=\mathbb{P}\left(e^{-Q}\leq x\right)=\mathbb{P}\left(-Q \leq \ln(x)\right)=\mathbb{P}\left(Q \geq -\ln(x)\right) \\
&= 1- \mathbb{P}\left(Q \leq - \ln(x)\right)=1-F_Q \left(-\ln(x)\right).
\end{align*}

By deriving the previous expression, we have:
\begin{equation*}
f_X(x)= f_Q (-\ln(x)) \frac{1}{x}= \frac{2}{\alpha} \left(\frac{-\ln(x)}{\alpha}\right)^2 \phi\left(\frac{-\ln(x)}{\alpha}\right) \frac{1}{x}= \frac{2}{\alpha x} \left(\frac{\ln(x)}{\alpha}\right)^2 \phi\left(\frac{\ln(x)}{\alpha}\right).   \quad \blacksquare
\end{equation*}

\end{document}